\documentstyle[12pt]{article}
\begin{document}
\begin{titlepage}
\vspace{10.0cm}
\title{{\bf Extended Inflation from Strings}
\thanks{Work partially supported by CICYT under
contract AEN90--0139.}}

\author{
{\bf J. Garc\'{\i}a--Bellido}\thanks{Supported by FPI Grant.
e--mail: bellido@iem.csic.es }\ \
and \  {\bf M. Quir\'{o}s}\thanks{
e--mail: imtma27@cc.csic.es and quiros@cernvm.} \\
Instituto de Estructura de la Materia\\
Serrano, 123\ \ \ E--28006\ \ Madrid.\ \ Spain}

\date{}
\maketitle
\vspace{1.5cm}
\def\baselinestretch{1.15}
\begin{abstract}

{We study the possibility of extended inflation in the effective
theory of gravity from strings compactified to four dimensions
and find that it strongly depends on the mechanism of
supersymmetry breaking. We consider a general class of
string--inspired models which are good candidates for successful
extended inflation. In particular, the $\omega$--problem of ordinary
extended inflation is automatically solved by the production of
only very small bubbles until the end of inflation. We find that
the inflaton field could belong either to the untwisted or to the
twisted massless sectors of the string spectrum, depending on
the supersymmetry breaking superpotential.}
\end{abstract}

\vskip-16cm
\rightline{{\bf IEM--FT--37/91}}
\rightline{{\bf April 1991}}
\vskip3in

\end{titlepage}

\newpage
\def\baselinestretch{1.25}

It is nowadays commonly believed \ by most cosmologists \ that the
inflationary paradigm may solve most of the problems of the
standard cosmological model. However, it is not yet clear how
the inflationary scenario can be successfully implemented. In
fact, the first proposed inflationary model (known as 'old'
inflation) \cite{G}, based on a first order phase transition, could
not provide a satisfactory explanation of how to get out from the
inflationary phase without disturbing the good properties of the
standard cosmological model \cite{G,GW}. The first models proposed
to solve this 'graceful exit' problem, known as 'new' inflation
\cite{NI}, with a second order phase
transition, were plagued with severe fine--tunings.
Soon after, a different solution without phase transition,
known as 'chaotic' inflation \cite{CI}, was proposed. (Chaotic
inflation has been recently shown to considerably soften the
fine--tuning problems of new inflation \cite{CEI}.)

Recently, La and Steinhardt \cite{EI} proposed an inflationary model,
known as 'extended' inflation, based again on a first order phase
transition, where the graceful exit problem was solved by using a
Jordan--Brans--Dicke \cite{JBD} theory of gravity with a scalar
field, instead of General Relativity \cite{WEI}. It was soon
realized that the anisotropy at decoupling produced by large
bubbles \cite{EW} made extended inflation incompatible with the
post--Newtonian bounds \cite{SCT} of General Relativity. This
desease could be cured either by using a more general
scalar--tensor theory of gravity \cite{MQ} or by means of a
cosmological constant with a runaway dependence on
the scalar field \cite{HKW}.

Most particle physicists believe that the theory of gravity
at low energies (General Relativity, scalar--tensor theories or
whatever) is an effective approximation of some fundamental
theory of quantum gravity at scales beyond the Planck mass ($M_P$).
The only reliable candidates for such a fundamental theory are
superstrings \cite{GSW}. They are known to describe gravity as a
low energy effective theory. It is therefore of
interest to know whether or not strings could lead to any kind
of cosmological inflation in the low energy effective theory.

The effective theory of superstrings exhibits three properties
that makes it a good candidate for some kind of extended
inflation. First of all, the scalar fields of the gravitational
multiplet (the dilaton and the moduli) are always coupled to the
curvature scalar of the four--dimensional
metric, in the same way as the scalar field in
scalar--tensor theories of gravity. Second, the dilaton and moduli
are also coupled to the matter sector giving rise to a scalar field
dependent potential. Finally, the existence of flat directions in the
potential follows from very general principles \cite{GSW}. In the
presence of supersymmetry breaking terms and a positive vacuum
energy, flat directions become runaway fields, and so are good
candidates for solving the anisotropy/post--Newtonian bounds
conundrum of extended inflation.

In a previous paper \cite{FT27} we studied the conformal properties
and cosmological solutions in the low energy effective theory
of gravity from closed strings compactified to four dimensions,
for different supersymmetric and non--supersymmetric string
scenarios, during the radiation and matter dominated eras.
In this paper we study the possibility of
extended inflation in string scenarios with spontaneously
broken supersymmetry. This problem has been recently
studied under some assumptions (in particular, constant values
for the moduli) with negative results \cite{LIN}.
However, we will show that the possibility of extended
inflation strongly relies on the mechanism of supersymmetry
breaking and find the conditions under which it could happen.
We will argue that a general necessary condition is the
existence of a positive 'metastable' minimum with some runaway
direction along the scalar fields.
This runaway direction should become flat at the true
minimum in order to solve the cosmological constant problem. In
this case, the same symmetry principle (if any) that could help
solving the cosmological constant problem, could also help
extended inflation.
A non--constant value of the moduli along the runaway
direction will help overcoming the problems found in
Ref.\cite{LIN}.

\vspace{1cm}

At string tree level, and keeping only linear terms in the
string tension $\alpha'$, the effective Lagrangian for the
dilaton $(\phi)$, the modulus $(\sigma)$ \footnote{We take the
diagonal direction in the space of moduli fields.}
and the matter fields
$(C_n)$, can be written in the Einstein frame as \footnote{We
will work hereafter, unless explicit mention, in units in which
$M_P\equiv 1$.} \cite{D4,DKL}
\begin{equation}
\label{L4W}
{\cal L}_{eff}= \sqrt{-\tilde{g}} \left[
\tilde{R} - \frac{1}{2}(\partial_\mu\phi)^2
- 6(\partial_\mu\sigma)^2  - \sum^3_{n=1} \frac{\alpha_n}{2^n}
 e^{-n(\sigma+\frac{1}{2}\phi)}\mid D_{\mu}C_n\mid^2
- V(\phi, \sigma, C_n, C_n^*) \right]
\end{equation}
where $C_1$ are untwisted matter fields, $C_2$ twisted
moduli (blowing up modes) and $C_3$ twisted matter fields,
and $\alpha_n$ are some positive constants ($\alpha_1=3$,
$\alpha_2=\alpha_3=1$).

For the purpose of this paper, in order to establish the
necessary conditions for extended inflation, it will be enough
to expand the potential $V$ in (\ref{L4W}) as
\begin{equation}
\label{Vm}
V(\phi, \sigma, C_n, C_n^*) = V_o(\phi,\sigma) +
\sum^3_{n=1} V_n(\phi,\sigma) \mid C_n\mid^2 + \ ...
\end{equation}
In fact, the Lagrangian (\ref{L4W}) and the potential (\ref{Vm})
can be put in the standard supergravity form \cite{CFG} by means
of the K\"{a}hler potential \cite{D4,DKL}
\begin{equation}
\label{Kp}
K = - \ln(S+S^*) - 3\ln(T+T^*) +
\sum^3_{n=1} \alpha_n(T+T^*)^{-n} \mid C_n\mid^2 +\ ...
\end{equation}
and the superpotential
\begin{equation}
\label{Wp}
W(S,T,C_n) = W_o(S,T) + \sum^3_{n=1} W_n(S,T)\ C_n^3 +\ ...
\end{equation}
where
\begin{equation}
\begin{array}{l}
\label{ST}
Re S = e^{3\sigma-\frac{1}{2}\phi} \vspace{2mm}\\
Re T = e^{\sigma+\frac{1}{2}\phi}
\end{array}
\end{equation}
It is important to stress that a superpotential $W_o$ different
from zero, necessary for supersymmetry breaking, and a
non--constant superpotential $W_n$ could be generated by
string non--perturbative effects. Also note that we are
consistently studying the Lagrangian along the (strong
CP--conserving) real directions \footnote{Notice
that a minimum along a different direction would just
amount to a field redefinition and so the general results of
this paper should remain valid.} ($Im S=Im T=0$).

The properties of the potential (\ref{Vm}), and in particular
its ability to produce extended inflation, will depend in
general on the form of the superpotential (\ref{Wp}). We will
first give some (by no means sufficient)
conditions on the potential (\ref{Vm}),
and their implications on the superpotential $W_o$, in order to
produce extended inflation:

a) We assume that the potential $V_o$ has a minimum along some field
direction, $e.g.$
\begin{equation}
\label{sb}
\sigma = - b \phi + d
\end{equation}
(with $b$ and $d$ some real parameters) and a runaway direction
along the orthogonal field\footnote{Otherwise $\phi$ and
$\sigma$ would be fixed to their vacuum expectation values and
no extended inflation could be present.
Since we are concerned in this paper with extended inflation from
strings, we will not consider the latter case.
On the other hand, the possibility of new inflation from strings
was studied some years ago and shown to require a huge amount of
fine--tuning \cite{ENQ}. Although these negative results are not general
enough to exclude other kinds of inflation based on General
Relativity ($e.g.$ chaotic) which could arise from string theories,
they make us search for inflation in more general theories of
gravity.}. This condition can be fulfilled
depending on the functional form of the superpotential. For
instance, if $W_o=W_o(X)$ with $X=S^\alpha T^{3\beta}$
($\alpha$ and $\beta$ real), then \footnote{The case $\alpha=0$,
$\beta=1/3$, giving $b=1/2$, has recently been considered
\cite{FIL} and shown to be consistent with target space modular
invariance. However, we will consider a more general case since
non--perturbative effects could break modular invariance \cite{GK}.}
\begin{equation}
\label{b}
b=\frac{3\beta-\alpha}{6(\alpha+\beta)}
\end{equation}
and the potential $V$ takes the form \cite{FIL}
\begin{equation}
\begin{array}{l}
\label{Von}
V_o=\frac{1}{16} e^{-6\sigma-\phi} \ v_o(\sigma+b\phi)
\vspace{2mm}\\
V_n=\frac{\alpha_n}{16\cdot 2^n} e^{-(6+n)\sigma-(1-\frac{n}{2})\phi}
\ v_n(\sigma+b\phi)
\end{array}
\end{equation}
where
\begin{equation}
\begin{array}{l}
\label{von}
v_o(\sigma+b\phi)=f^2_\alpha + 3f^2_\beta - 3f^2_o
\vspace{2mm}\\
v_n(\sigma+b\phi)=f^2_\alpha + (3-n)f^2_\beta - 2f^2_o
\end{array}
\end{equation}
with
\begin{equation}
\label{fl}
f_\lambda (\sigma+b\phi)\equiv W_o-2\lambda X\frac{\partial W_o}
{\partial X}   \ .
\end{equation}
The minimization of $v_o$ in (\ref{Von}) should provide the
condition (\ref{sb}).

Notice that condition (\ref{sb}) is not essential for
extended inflation. It is just a simplifying assumption where
one direction in the $(\sigma,\phi)$ configuration space is
fixed to its vacuum expectation value and so the
remaining theory of
gravity has only one scalar field. However, more general
situations suitable for extended inflation are thinkable. For
instance, the case where both $\sigma$ and $\phi$ are runaway
directions (no field is fixed to its vacuum
expectation value) can be easily realized in many
models. In particular, in the simple case where $W_o=$ constant.
(A constant superpotential can be triggered by the vacuum
expectation value of some field.) In this case, $\ v_o\ =\ \mid
W_o\mid^2\ $ and $\ v_n\ =\ (2-n)\\
\mid W_o\mid^2$.

b) There should be a positive cosmological constant at the
minimum (\ref{sb}), $i.e.$
\begin{equation}
\label{cc}
v_o(d)>0  \ .
\end{equation}
In particular, this implies that supersymmetry is broken at the
minimum (\ref{sb}) in such a way that
\begin{equation}
\label{cm}
f^2_\alpha(d) + 3f^2_\beta(d) > 3f^2_o(d)   \ .
\end{equation}
In the case $W_o=$ constant, condition (\ref{cc}) is
automatically satisfied.

c) The last condition is that the minimum (\ref{sb}) is required
to be stable along the inflaton field direction $C_n$, $i.e.$
\begin{equation}
\label{cd}
v_n(d)>0
\end{equation}
or
\begin{equation}
\label{cn}
f^2_\alpha(d) + (3-n)f^2_\beta(d) > 2f^2_o(d)
\end{equation}
where $n$ is the sector to which the inflaton belongs.
In this way, the inflaton potential can trigger a first
order phase transition from the false vacuum at $C_n=0$ to the
true physical vacuum at $C_n\neq0$, which we assume to
correspond to a zero cosmological constant \footnote{Of course
this would impose extra conditions on the total
superpotential $W$, which we will not study here.}.

In the simple case of $W_o=$ constant, condition
(\ref{cd}) is always satisfied for $n=1$ (untwisted matter
sector) but never satisfied for $n=3$ (twisted matter sector).
For $n=2$ (blowing--up modes) $v_2\equiv 0$ and so the stability
along the inflaton direction $C_2$ would rely upon higher order
derivatives of the potential and therefore upon the
superpotential $W_2$.

\vspace{1cm}

In what follows we will assume that conditions (\ref{sb}),
(\ref{cc}) and (\ref{cd}) hold and therefore will write the
Lagrangian (\ref{L4W}) for $\phi$ and the inflaton
field $C_n$ as
\begin{equation}
\label{L4}
{\cal L}_{eff}= \sqrt{-\tilde{g}} \left[
\tilde{R} - (6b^2+\frac{1}{2})(\partial_\mu\phi)^2
- e^{-n(\frac{1}{2}-b)\phi}\mid D_\mu C_n\mid^2
-\ e^{-(1-6b)\phi}\rho_o + ...\right]
\end{equation}
where $\rho_o$ is a constant energy density, we have used
Eq.(\ref{sb}) and absorbed all constant coefficients in the
definition of $C_n$.
Notice that the energy density $\rho(\phi)$ in (\ref{L4}) is
proportional to $m^2_{3/2}$, the scale of supersymmetry
breaking (the gravitino mass), as expected,
\begin{equation}
\label{m3/2}
m^2_{3/2}\propto e^{-(1-6b)\phi} \mid W_o \mid^2 \ .
\end{equation}
The mass of the observable fields at the true vacuum depends on
the global structure of the potential $V$ in (\ref{L4W}), which
is very poorly known in most cases. In fact, it depends on the
total structure
of the K\"ahler potential (\ref{Kp}) and the superpotential
(\ref{Wp}), which could in turn depend on non--perturbative
effects at high energy scales (string effects) and/or at low
energy scales (QCD condensates, ...). We will assume for the
masses a simple dependence
\begin{equation}
\label{m2}
\tilde{m}^2 \sim e^{-a\phi}\ m^2_o
\end{equation}
where $m_o$ is a constant mass and $a$ is a real coefficient
parametrizing our ignorance on the details of supersymmetry
breaking in string theory and the low energy
non--perturbative effects.
The case of constant masses ($a=0$), considered in the analysis
of Ref.\cite{LIN}, is particularly interesting and will be
commented later on.

Under a conformal redefinition \cite{DGG,KKO,LIN,FT27} of the metric
\begin{equation}
\label{DG}
\tilde{g}_{\mu \nu} = e^{c\phi}\ g_{\mu \nu}
\end{equation}
\begin{equation}
\label{RR}
\tilde{R}=e^{-c\phi}\left[ R - c(D-1) D^{2}\phi-\frac{1}{4}
c^2(D-1)(D-2) g^{\mu\nu}\partial_{\mu}\phi \partial_{\nu}\phi
\right] \ ,
\end{equation}
the masses transform as
\begin{equation}
\label{mp}
m^2=e^{c\phi}\ \tilde{m}^2  \ .
\end{equation}

It is therefore convenient to make the conformal redefinition of
$g_{\mu\nu}$ (\ref{DG}, \ref{RR}) with parameter $c=a$ such that
the mass of the observable fields, see Eqs.(\ref{m2}, \ref{mp}),
become $\phi$--independent \cite{FT27}. Then (\ref{L4}) can
be written as \footnote{Recall that under a conformal
redefinition of the Robertson--Walker metric, the scale factor
and the time variable transform as
$\tilde{a}(\tilde{t})=\Phi(t)^{1/2} a(t)$ and
$d\tilde{t}=\Phi(t)^{1/2} dt$ respectively.}
\begin{equation}
\label{L4P}
{\cal L}= \sqrt{-g}\left[\Phi R -
\frac{\omega}{\Phi}(\partial_\mu\Phi)^2
- \frac{1}{2}\Phi^{1-\beta'}(\partial_\mu C_n)^2
- \Phi^{2(1-\beta)}\rho_o \right]
\end{equation}
where
\begin{equation}
\label{F}
\Phi= e^{a\phi}
\end{equation}
and the parameters $\omega$, $\beta$ and $\beta'$ in (\ref{L4P}) are
defined as functions of $a$ and $b$ as
\begin{equation}
\label{OM}
2\omega+3=\frac{1+12b^2}{a^2}
\end{equation}
\begin{equation}
\label{beta}
\beta=\frac{1-6b}{2a}
\end{equation}
\begin{equation}
\label{beta'}
\beta'= n\left(\frac{1-2b}{2a}\right)  \ .
\end{equation}

Written in terms of a Robertson--Walker metric,
$\Phi(t)$ is a dimensionless
scalar related to the variation of the Plank mass
\begin{equation}
\label{Phi}
\Phi(t) = \frac{M^2_P(t)}{M^2_P}
\end{equation}
where $M_P^2$ stands for $M_P^2(t_o)\equiv 1/G_N$
($t_o$ is the present age of the universe),
and we assume $\Phi(t_e)\simeq 1$ at the end of inflation.
We can also define the scales $M$ and $m_P$ through
\begin{equation}
\label{M}
\rho(0) = \Phi(0)^{2(1-\beta)} \rho_o \equiv M^4
\end{equation}
\begin{equation}
\label{Ph0}
\Phi(0) \equiv \frac{m^2_P}{M^2_P}   \ .
\end{equation}

The equations of motion then read
\begin{equation}
\label{SPF}
\begin{array}{c}
{\displaystyle
\left(\frac{\dot{a}}{a}
\right)^2+\frac{k}{a^2}=
\frac{\rho_o}{6}\Phi^{1-2\beta} + \frac{\omega}{6}
\left(\frac{\dot{\Phi}}{\Phi} \right)^2-\frac{\dot{a}}{a}
\frac{\dot{\Phi}}{\Phi} }\vspace{2mm}\\
{\displaystyle
\ddot{\Phi} + 3\frac{\dot{a}}{a}\dot{\Phi}
=\frac{2\beta}{2\omega+3} \rho_o \Phi^{2(1-\beta)} }
\end{array}
\end{equation}
with solutions for $k=0$ \cite{HKW}
\begin{equation}
\label{SS}
\begin{array}{l}
a(t)=a(0) (1+Bt)^p, \hspace{2cm} {\displaystyle p=
\frac{2\omega + 3 -2\beta}{2\beta(2\beta-1)} } \vspace{2mm}  \\
\Phi(t)=\Phi(0) (1+Bt)^q, \hspace{2cm}
{\displaystyle q=\frac{2}{2\beta-1} }
\end{array}
\end{equation}
where
\begin{equation}
\label{B}
B^2 = \frac{2\beta^2(2\beta-1)^2\rho_o \Phi(0)^{1-2\beta}}
{(2\omega+3)(6\omega+9-4\beta)} =
\frac{2\beta^2(2\beta-1)^2\ M_P^2}{(2\omega+3)(6\omega+9-4\beta)}
\left(\frac{M}{M_P}\right)^4\left(\frac{M_P}{m_P}\right)^2    \ .
\vspace{2mm}
\end{equation}

We now raise the question of the sufficient conditions for
extended inflation and whether or not
a 'gracefull exit' can be achieved.

First of all, we require that quantum gravity effects be
negligible. In other words, that the kinetic energy due to de
Sitter fluctuations (maximal at beginning of inflation \cite{ST,EI})
be less than $\rho$, see Eqs.(\ref{M}--\ref{SPF}), $i.e.$
${\displaystyle H^4(0) \approx M^8
\left(\frac{M_P}{m_P}\right)^4 < \rho(0) }$. This gives the constraint
\begin{equation}
\label{mP}
m_P > M   \ .
\end{equation}

We are assuming that the universe at $T_c$ goes through a first
order phase transition in which the high-temperature phase remains
metastable down to $T=0$ \cite{GW}, where bubble nucleation is
dominated by quantum mechanical tunneling \cite{COL}. Bubbles are
assumed to be formed with zero radius at $t_B$ and then expand at
the speed of light. A bubble radius at a later time $t>t_B$ is
given by
\begin{equation}
\label{rad}
r(t,t_B)=\int^t_{t_B}\frac{dt'}{a(t')}   \ .
\end{equation}
We now define the asymptotic radius of a bubble nucleated at $t$
as
\begin{equation}
\label{ras}
r_{as}(t)=\int^\infty_t \frac{dt'}{a(t')}
= \frac{p}{p-1}\ \frac{1}{a(t) H(t)}
\end{equation}
where $H(t)$ is the Hubble expansion factor
\begin{equation}
\label{H}
H(t)=pB \left(\frac{\Phi(0)}{\Phi(t)}\right)^{\beta-1/2} =
pB \left(\frac{a(0)}{a(t)}\right)^{1/p}   \ .
\end{equation}

The end of inflation is determined by the competition
between the bubble nucleation rate and the cosmic expansion
rate. The dimensionless parameter which controls the percolation
of the phase transition can be computed as \cite{GW}
\begin{equation}
\label{eps}
\epsilon(t)=\int^t_{t_B} dt' \lambda(t') a^3(t') \frac{4\pi}{3}
r^3(t,t') \simeq \frac{\lambda(t)}{H^4(t)} \hspace{2cm} (p\gg 1)
\end{equation}
where $\lambda(t)$ is the bubble nucleation rate per unit volume.
In our model, $\lambda(t)$ is time dependent since the energy
density which drives inflation is itself time dependent through
$\Phi(t)$, see Eq.(\ref{L4P}). Holman {\it et al.} \cite{HKVW}
compute this dependence to be
\begin{equation}
\label{lam}
\lambda(t)=\lambda_o \Phi(t)^{2(1-\beta')}
e^{-B_o\left[\Phi(t)^{2(\beta-\beta')}-1\right]}
\end{equation}
where \footnote{In ordinary JBD theories and GR, this rate is
essentially time independent and given by $\lambda_o$.}
$\lambda_o=Ae^{-B_o}$. $B_o$ is the Euclidean bounce action
\cite{COL,GW}, which depends on the inflaton
potential and can acquire very big values $O(10^2)$, while the
prefactor $A$ is of order one and has dimensions of $T_c^4$,
where $T_c\sim M$ is the mass scale of the phase transition.

The epsilon parameter can then be written as
\begin{equation}
\begin{array}{rl}
\label{eps0}
\epsilon(t)&=\epsilon_o \ \Phi(0)^{2(1-2\beta)}
\Phi(t)^{2(2\beta-\beta')}
e^{-B_o\left[\Phi(t)^{2(\beta-\beta')}-1\right]} \vspace{2mm}\\
&=\epsilon(t_e) \Phi(t)^{2(2\beta-\beta')}
e^{-B_o\left[\Phi(t)^{2(\beta-\beta')}-1\right]}
\end{array}
\end{equation}
where ${\displaystyle \epsilon_o\equiv\frac{\lambda_o}{H^4(0)} }$
is the usual parameter considered in the literature.

A measure of the progress of the transition is the fraction of
space which remains in the high temperature phase ('false vacuum'),
$p(t)=e^{-\epsilon(t)}$. We need a very small epsilon
parameter at the beginning of inflation which increases very
quickly to above a critical value, thus allowing for percolation
of the low temperature phase ('true vacuum').
Therefore we require
\begin{equation}
\label{conE}
\epsilon(t_e)=\epsilon_o\ \Phi(0)^{2(1-2\beta)} \ > \epsilon_{cr}
\end{equation}
where ${\displaystyle 1.1\times 10^{-6}<\epsilon_{cr}<\frac{3}{4\pi} }$
was computed
in Ref.\cite{GW} for solving the 'gracefull exit' problem. Thus
\begin{equation}
\label{e0}
\epsilon_o > \left(\frac{M}{M_P}\right)^{4(2\beta-1)} \epsilon_{cr}
\end{equation}
which gives ample room for very small values of $\epsilon_o$,
provided that $2\beta>1$ (which is anyhow necessary for an increasing
$\Phi(t)$). We must be sure, however, that $\epsilon(t)$ is increasing,
that is
\begin{equation}
\label{deps}
\frac{\dot{\epsilon}(t)}{\epsilon(t)} = 2(\beta'-\beta)
\ \frac{\dot{\Phi}(t)}{\Phi(t)} \left[B_o\Phi(t)^{-2(\beta'-\beta)}
-\frac{\beta'-2\beta}{\beta'-\beta}\right] > 0
\vspace{1mm}
\end{equation}
which is satisfied for
\begin{equation}
\label{cde}
\beta' > \beta
\end{equation}
and
\begin{equation}
\label{cds}
B_o\Phi(t)^{-2(\beta'-\beta)} >
\frac{\beta'-2\beta}{\beta'-\beta} \ .
\end{equation}
This condition is very easily satisfied as we will see.

We are now ready to analyze our string model for inflation,
see Eq.(\ref{L4P}). The peculiarity of this model is the fact
that $\omega$, $\beta$ and $\beta'$ are
not independent but determined by the string effective
action, see Eqs.(\ref{OM}--\ref{beta'}).
This dependence corresponds to the conformal redefinition of the
metric tensor for which observable matter have constant masses, as
discussed above. We will now impose further constraints on our model.

A necessary condition for inflation is that
${\displaystyle \frac{\ddot{a}}{a}>0  }$, or $p>1$, which becomes
\begin{equation}
\label{b1/2}
0 < b < \frac{1}{2}   \ .
\end{equation}
We must also impose that $\Phi(t)$ increases,
which gives the condition
\begin{equation}
\label{a6b}
a < 1-6b   \ .
\end{equation}
The condition that $\epsilon(t)$ increases, see
Eqs.(\ref{cde}, \ref{cds}), then becomes
\begin{equation}
\label{a0}
a \geq 0
\end{equation}
\begin{equation}
\label{B0}
B_o > 1
\end{equation}
which are both sufficient conditions for all values of $n$, see
Eq.(\ref{beta'}).

Assuming $N$ orders of magnitude increase in the scale factor,
\begin{equation}
\label{10N}
10^N=\frac{a(t_e)}{a(0)}=\left(\frac{\Phi(t_e)}{\Phi(0)}\right)^{p/q}
< \left(\frac{M_P}{M}\right)^{p/q}
\end{equation}
imposes the constraint
\begin{equation}
\label{abM}
a < \left(\frac{1+12b^2}{1-6b}\right)\
\frac{\log \left(\frac{M_P}{M}\right)}
{N+\log\left(\frac{M_P}{M}\right)}   \ .
\end{equation}

Furthermore, in order to solve the horizon problem we need
sufficient inflation such that
\footnote{We use here the notation $H_o\equiv H(t_o)$,
$a_o\equiv a(t_o)$ and $T_o\equiv T(t_o)$.}
${\displaystyle d_{H_o}<d_{H(0)}\frac{a_o}{a(0)} }$.
However, since $H(t)\sim t^{-1}\sim T^2$ and $a T\sim$ constant
during the post--inflationary period, and assuming 'good
reheating' for recovering all the latent energy density of the
phase transition ($T_e\equiv T(t_e)\sim T_c\sim M$), we obtain the
condition
\begin{equation}
\label{N}
N > \frac{p}{p-1} \log \left(\frac{M}{T_o}\right)  \ .
\end{equation}
Therefore, the required number of orders of magnitude of
inflation depends
crucially on the energy scale of the phase transition $M$.

Inflation must occur after the production of monopoles or any
other topological defects whose densities might affect
cosmology. For the same reason, the universe must also reheat
before baryogenesis. These conditions place the constraint
$10^2$ GeV $<M<$ $10^{18}$ GeV \cite{EI}.

However, solving the horizon and monopole problems is not
enough. We must be sure that the phase transition ends and
that all the bubbles percolate without disturbing too much the
isotropy and homogeneity of the cosmic background radiation.
Therefore, we expect that the volume fraction contained in
bubbles with radius greater than a given comoving radius
$r=r(t_e,t)$ at the end of inflation be less than $10^{-n}$ at a
temperature $T$:
\begin{equation}
\label{V}
{\cal V}_>(r,t_e) = 1-p(t) \simeq \epsilon(t) = \epsilon(t_e)
\ \left(\frac{T}{M}\right)^\delta \ e^{-B_o \left[
\left(\frac{M}{T}\right)^{\delta'}-1\right]} \ < \ 10^{-n}
\end{equation}
where we have used
\begin{equation}
\label{del}
\Phi(t)^{2(2\beta-\beta')}=\left(\frac{r_o}{r}\right)^\delta
\simeq \left(\frac{T}{M}\right)^\delta
\vspace{2mm}
\end{equation}
where $r_o\equiv r_{as}(t_e)$ is the asymptotic
radius of a bubble nucleated at the end of inflation,
$\ {\displaystyle \delta\equiv\frac{8\beta(2\beta-\beta')}
{2\omega+3-4\beta^2} }\ $
and $\ {\displaystyle \delta'\equiv\frac{8\beta(\beta'-\beta)}
{2\omega+3-4\beta^2} > 0}\ $. In particular, for the cosmic
background radiation, we require that \cite{EW}
$n\simeq 5$ at $T\simeq 1$ eV in (\ref{V}).
This condition is trivially satisfied thanks to the exponential,
using ${\displaystyle \frac{M}{T}>10^{11} }$ and condition (\ref{B0}).
In this way, the extended inflation problem of anisotropy at
decoupling produced by large bubbles is successfully solved in
this kind of models \footnote{In fact, this solution was proposed on
general grounds in Ref.\cite{LIN}.}.

We still have to be sure of reestablishing a common
Robertson--Walker frame in all the bubble clusters that will
coalesce to form our universe. There must be some way to
remember the original (pre--bubble--nucleation) coordinates; such
a record can be found in the evolution of $a(t)$ or $\Phi(t)$.
Since constant $H(t)$ corresponds in General Relativity to a de Sitter
universe with no distinguished frame, we must require sufficient
variations of this quantity, $e.g.$ $m$ orders of magnitude
in $H(t)$ \cite{EW,HKW}. In particular, we expect that homogeneity
and isotropy must hold by the time of nucleosynthesis
($T_{NS}\simeq 1$ MeV, $m\simeq 1$), thus
\begin{equation}
\label{Hm}
\frac{H(t)}{H(t_e)}=\left(\frac{r+r_o}{r_o}\right)^{\frac{1}{p-1}}
\simeq\left(\frac{M}{T}\right)^{\frac{1}{p-1}}\ > \ 10^m
\end{equation}
corresponding to
\begin{equation}
\label{ppo}
p < 1 + \log\left(\frac{M}{T_{NS}}\right) \equiv p_o
\end{equation}
which is an explicit bound on the power of the scale factor
and gives an extra condition on our parameters
\begin{equation}
\label{apo}
a < \frac{p_o}{p_o-1} (1-6b) - \frac{1}{p_o-1}
\left(\frac{1+12b^2}{1-6b}\right)  \ .
\end{equation}

Furthermore, quantum fluctations of the scalar field $\Phi$
create a spectrum of adiabatic fluctuations, which can be
computed for power--law solutions in the Einstein frame \cite{LM}
\begin{equation}
\label{rpl}
\frac{\delta\tilde{\rho}}{\tilde{\rho}}\simeq
\frac{\tilde{H}^2}{\pi\dot{\phi}}\simeq
\frac{\tilde{p}^{3/2}}{\pi}\cdot\frac{1}{\tilde{t}}
\end{equation}
and must be bounded in the conformal frame ($\rho=\Phi^2\tilde{\rho}$)
to be compatible with the observed density perturbations \cite{KKO}
\begin{equation}
\label{drho}
\frac{\delta\rho}{\rho}\simeq\left(\frac{M}{M_P}\right)^2 \tilde{p}
\ \ k^{-\frac{1}{\tilde{p}-1}} < 10^{-4} \hspace{2cm}  (\tilde{p}\gg 1)
\end{equation}
where $k$ is the dimensionless physical scale of reentering
perturbations and $\tilde{p}$ is the power of the scale
factor in the Einstein frame ${\displaystyle \left(
\tilde{a}(\tilde{t})\sim \tilde{t}^{\ \tilde{p}},\ \ \tilde{p}=
\frac{2\omega+3}{4\beta^2}\right) }$.
This imposes a very mild constraint on $M$
\begin{equation}
\label{MMp}
M < \left(\frac{1-6b}{\sqrt{1+12b^2}}\right)\ 10^{-2} M_P < 10^{-2} M_P \ .
\end{equation}

Finally, the most stringent bounds will come from the
post--Newtonian experiments of time delay \cite{VIK,SCT}
and the nucleosynthesis bound \cite{FT26} on the $\omega$
parameter
\begin{equation}
\label{pN}
2\omega + 3 > 500 \hspace{2cm} (2\sigma)
\end{equation}
which gives a very strong constraint on our parameters
\begin{equation}
\label{aom}
a < \sqrt{\frac{1+12b^2}{500}}    \ .
\end{equation}

It is interesting to notice that the anisotropy of the cosmic
background radiation, which was the main problem for extended
inflation, does not impose any significant bound on our
model, see Eq.(\ref{V}). The most stringent bound comes from the
post--Newtonian experiments and nucleosynthesis bound, see
Eq.(\ref{aom}), which constraint the parameter $a$. On the
other hand, the strongest constraint on $b$ comes from the
isotropy and homogeneity at nucleosynthesis, see
Eqs.(\ref{ppo}, \ref{apo}).

Most of the previous bounds depend on the energy scale $M$ of
the phase transition.  We have studied those bounds for two typical
values of $M$. For a phase transition driven by phenomenological
supersymmetry breaking ($m_{3/2}\simeq 1$ TeV) we have
$M=(m_{3/2} M_P)^{1/2}\simeq 10^{11}$ GeV,
while for the usual grand unified theory we take $M=10^{16}$ GeV.
The inflationary scenario is characterized by two parameters,
the power $p$ of the scale factor and the number $N$ of orders of
magnitude increase during inflation. Both parameters depend
on the energy scale of the phase transition, see
Eqs.(\ref{N}, \ref{ppo}). For $M=10^{11}$ GeV we have $p<16$ and
$N>26$, while for $M=10^{16}$ GeV, $p<21$ and $N>31$ to solve
the horizon problem without disturbing the isotropy and
homogeneity at nucleosunthesis. The actual value of $N$ depends
on the parameters of the theory. Using the bounds (\ref{apo}),
(\ref{MMp}) and (\ref{aom}) we obtain $N>45$, which widely
solves the flatness problem.

In Fig.1 we show the region in parameter space $(a, b)$ allowed
by all the inflationary conditions, for $M=10^{11}$ GeV and
$M=10^{16}$ GeV (dashed and dotted curves respectively). The
allowed region is
the one below the curves. The condition associated with
neglecting the quantum gravity effects (\ref{mP}, \ref{abM})
strongly depends on the energy scale of the phase transition, as
expected, and as we can see from the lines labelled QG. Other
conditions depend slightly on $M$, like those associated with
reestablishing the isotropic and homogeneous Robertson--Walker
frame (\ref{Hm}, \ref{apo}), and labelled RW in Fig.1 . Finally,
there are those conditions which do not depend at all on the
energy scale of the phase transition, like the post--Newtonian
bounds (\ref{pN}, \ref{aom}) and the condition (\ref{a6b})
that $\Phi$ increases from $m_P$ to $M_P$, labelled
pN--NS and $\Phi$ respectively.
However, as we can see from Fig.1, the final allowed region in
parameter space does not depend much on the scale $M$ since
it is bounded by the post--Newtonian and nucleosynthesis bound and
the isotropy and homogeneity condition at nucleosynthesis.

As we can see from Fig.1, the case of constant observable masses
($a=0$) is consistent with all inflationary and post--Newtonian
bounds. This can be easily obtained  by taking the limit
$a\rightarrow0$ in our explicit solution (\ref{SS}), which
corresponds to
\begin{equation}
\label{sol}
\begin{array}{l}
a(t) \sim t^{\ \frac{1+12b^2}{(1-6b)^2}} \\
\Phi(t) \sim 1
\end{array}
\end{equation}
On the other hand the direction $b=0$, see Eq.(\ref{sb}), is
incompatible with the
necessary condition for inflation, Eq.(\ref{b1/2}), and
corresponds to the case of constant moduli. We thus agree with
the negative results found in Ref.\cite{LIN}.

In conclusion, we have studied in this paper the general
conditions under which the effective theory of gravity from
strings compactified to four dimensions could lead to extended
inflation. We have found that a necessary condition is the
existence of runaway directions in the space of fields coupled
to the curvature scalar (dilaton and moduli fields).
However, the existence of runaway directions is a usual feature
of the effective theory of strings (through classical invariance
arguments \cite{GSW}). It is satisfied for many supersymmetry
breaking potentials. In the simplest case of supersymmetry
breaking, a constant $W_o$ superpotential in (\ref{Wp}), all
moduli and the dilaton are runaway fields with a positive
potential ($\sim\mid W_o\mid^2$) and extended inflation may
follow. However, to simplify the study of the equations of
motion, we have assumed just one runaway direction and
parametrized it by a real parameter $b$. This is
just a simplifying hypothesis since extended inflation might
occur under much more general circumstances.

A second necessary condition for extended inflation
is the existence of a metastable
minimum along some (matter) inflaton field. This condition is
necessary to enforce a first order phase transition.
It also depends on the particular structure of the
supersymmetry breaking superpotential, but this condition (see
Eq.(\ref{cd})) is easily satisfied in many models. For instance,
in the simple case $W_o=$ constant, it holds when the inflaton
belongs to the untwisted sector ($n=1$), and does not hold
if it belongs to the twisted sector ($n=3$). The case of
the inflaton as a blowing--up mode ($n=2$) would require the
precise knowledge of the total superpotential.

Third, we assume a simple behaviour of the mass of observable
fields on the runaway direction, and parametrize this behaviour
with a real parameter $a$. (The case of constant masses
corresponds to $a=0$.) We make a conformal redefinition of the
metric in order to go into the 'physical' frame, where the masses of
the observable fields are constant. Of course, if the functional
dependence of masses were more complicated, we would have needed
a more general conformal transformation and the theory would
look different, in particular it would have a non--constant
$\omega$ parameter. However, we should stress here that a
conformal redefinition is not essential since Physics cannot
depend on it. In other words, we could redefine the physical
scale factor by taking its ratio with respect to the Compton
wavelength \cite{HKWW} which is then manifestly
independent of the conformal transformation \cite{LIN,FT27}.

Finally, we have imposed all the conditions for successful
extended inflation on the solution of our model and found an
allowed region in parameter space ($a,b$). Our results are
summarized in Fig.1. The direction $b=0$ (the region of constant
moduli) is excluded from the allowed region,
while $a=0$ (the case of constant masses) is
inside the permitted region and therefore consistent with all
experimental bounds. Notice that our model successfully solves
the $\omega$--problem of extended inflation (namely, that the
condition of isotropy of the cosmic background radiation at
decoupling is in conflict with the post--Newtonian bounds on
$\omega$) by producing very small bubbles until the end of
inflation when the epsilon parameter increases exponentially up to
the critical value.

\section*{Acknowledgements}
One of us (J.G.--B.) would like to thank M. Cveti\v{c}, B. Ovrut
and P. Steinhardt for valuable discussions, and the Theoretical
Physics Department of Pennsylvania University for the
hospitality given to him. The other (M.Q.) thanks F.
Quevedo for discussions and the T--8 Division of Los Alamos
National Laboratory for its warm hospitality.

\newpage
\section*{Figure Captions}
\begin{description}

\item[Fig.1]
Plot of the region in parameter space $(a, b)$ allowed by the
inflationary conditions in the text. The solid lines represent
those bounds
which do not depend on the scale $M$ of the phase transition.
The dashed curve correspond to the bounds associated with the
scale $M=10^{11}$ GeV and the dotted curve to those related to
$M=10^{16}$ GeV. The allowed region is the one below the curves.
The border $b=0$ is excluded from it.
We have labelled the curves as follows: QG corresponds to the
condition associated to neglecting quantum gravity effects,
$\Phi$ corresponds to the condition for an increasing scalar
field, RW corresponds to the bound on isotropy and
homogeneity at the time of nucleosynthesis and pN--NS
corresponds to the bounds from post--Newtonian experiments and
the nucleosynthesis bound on $\omega$.

\end{description}
\end{document}